# Kinetics of the Initial Stage of Immunoagglutination Studied with the Scanning Flow Cytometer

*Ivan V. Surovtsev[1], Maxim A. Yurkin[2], Alexander N. Shvalov[1], Vyacheslav M. Nekrasov[2], Galina F. Sivolobova[1], Antonina A. Grazhdantseva[1], Valeri P. Maltsev[2,3], and Andrey V. Chernyshev[2,3,*]*

[1] State Research Center for Virology and Biotechnology VECTOR
Koltsovo, Novosibirsk region 630559, Russia
[2] Institute of Chemical Kinetics and Combustion,
Institutskaya 3, Novosibirsk 630090, Russia
[3] Novosibirsk State University, Pirogova 2, Novosibirsk 630090, Russia

**ABSTRACT**
The use of a scanning flow cytometer (SFC) to study the evolution of monomers, dimers and higher multimers of latex particles at the initial stage of the immunoagglutination is described. The SFC can measure the light-scattering pattern (indicatrix) of an individual particle over an angular range of 10-60º. A comparison of the experimentally measured and theoretically calculated indicatrices allows one to discriminate different types of latex particles (i.e. monomers, dimers, etc.) and, therefore, to study the evolution of immunoagglutination process. Validity of the approach was verified by simultaneous measurements of light-scattering patterns and fluorescence from individual polymer particles. Immunoagglutination was initiated by mixing BSA-covered latex particles (of 1.8 μm in diameter) with anti-BSA IgG. The analysis of experimental data was performed on the basis of a mathematical model of diffusion-limited immunoagglutination aggregation with a steric factor. The steric factor was determined by the size and the number of binding sites on the surface of a latex particle. The obtained data are in good agreement with the proposed mathematical modeling.

**Key words:** immunoassay, latex particles, aggregation kinetics, scanning flow cytometer, light scattering indicatrix

**INTRODUCTION**

Agglutination is one of the oldest serological reactions that results in clumping of a cell suspension by a specific antibody directed against an antigen [1]. Nowadays this phenomenon is widely used in different applications. Although there are a number of high sensitive and resourceful techniques and methods, for example PCR, ELISA, that have been developed for diagnostics purposes, the interest to rapid, simple and inexpensive tests still remains [2]. Antigen-coated latex is used for quantifying antibody concentration in a specimen (passive agglutination), and antibody-coated latex for antigen concentration (reversed passive agglutination) [3]. The enhanced latex particle immunoagglutination test is widely used in the diagnostics of various infections and to detect biomarkers or some chemical compounds in biological fluids [4-6]. On the other hand, agglutination takes place in different biological processes, such as platelets or red blood cell aggregation [5]. The interest to agglutination is also supported by the general study of coagulating systems.

---

[*] Correspondence to: A. V. Chernyshev, Institute of Chemical Kinetics and Combustion, Institutskaya 3, Novosibirsk 630090, Russia. E-mail: chern@ns.kinetics.nsc.ru. Fax: 7-3832-342350. Tel.: 7-3832-333240.

Various techniques for quantitative immunoagglutination assays have been proposed that rely on turbidimetric methods [5], ultrasound instruments [6], electrophoretic mobility [7], centrifugation [8] and light scattering [9-15]. Experimental methods based on light scattering techniques have been used for measurement of colloidal dispersions [9-11] and single-particles [12-14].

Flow cytometry [15] is one of the most powerful and widely spread technologies for the measurement of single particles. In 1985 Cohen and coworkers developed the instrument based on light scattering and flow system, which allows the investigation of particles aggregation, if the particle size is small compared with the wavelength [12-14]. But the evaluation of the particle size and shape (and other parameters) from the light scattering measurements of the particles, which are big compared with the wavelength remains a problem.

Over the past decade the Scanning Flow Cytometry (SFC) has been developed for advanced characterization of single particles [16-21]. The advantage of the SFC is based on the measurement of the angular dependency of light scattered by an individual particle, i. e. its scattering pattern (indicatrix). The indicatrix depends on size, refractive index, shape and morphological structure of the particle. The SFC technology provides more detailed information about a single particle than the ordinary flow cytometry, since the indicatrix brings much more information than the integrated intensities of light scattered in forward and side angular ranges. It is believed that one of the most promising and attractive fields of application of the SFC technology is the characterization of non-spherical particles. Since clusters of two, three, and forth spheres arising from single microspheres are non-spherical, the use of SFC for quantitative study of immunoagglutination is expected to be very informative.

In the present work we first applied the SFC for the study of the evolution of monomers, dimers and higher multimers during initial stage of immunoagglutination. In order to verify the proposed method, we compared the results of particle discrimination obtained from light scattering indicatrix and fluorescence. Immunoagglutination was caused by the reaction of BSA antigens immobilized on the particles surfaces with rabbit anti-BSA antibodies dissolved in the media. A mathematical model of the evolution of monomers, dimers and higher miltimers at the initial stage of the latex agglutination has been applied, in order to treat the experimental data.

**MATERIALS and METHODS**
*Latex particles and antibodies*

We used carboxylated fluorescent latex microspheres of 1.8 μm in diameter from Polysciences (USA, PA). Carboxylate-modified particles have pendent carboxylic acids that make them suitable for covalent coupling with proteins or other amine-containing biomolecules. Two sets of latexes were employed in the experiments: uncovered particles and particles covered with bovine serum albumin (BSA, Sigma).

In order to activate surface carboxyl groups on the particles and to attach BSA to the particles surface covalently, we employed the wide spread method using water-soluble carbodiimide, namely 1-ethyl-3-(3-dimethylaminopropyl)carbodiimide (EDAC). The procedure is a simple one-step method of coupling of avidin, streptavidin,

BSA and immunoglobulins of different origins. After washing procedure latex particles covered with BSA were resuspended in 50 mM PBS (pH=7.4). Upon addition of the 2 mM sodium azide as preservative covered latex particles had been stored in a dark place at 4$^o$C without freezing before the experiments. Concentration of BSA-covered latex particles was $6.0 \cdot 10^8$ cm$^{-3}$ prior to the experiments.

Anti BSA polyclonal rabbit immunoglobulins of the class G (IgG) were kindly granted by Dr. G. M. Ignatiev (SRC VB VECTOR). The antibodies were used to initiate agglutination process in our study by adding them to the particles, which carried corresponding antigen. The IgG concentration was determined by Louri [22] method and was 1.6 μM prior to mixing.

*Experimental setup*

In this work, we used the SFC described elsewhere [18-21]. It should be noted that the main differences of the SFC from an ordinary flow cytometer are as follow: 1) the laser beam (Ar laser, 441 nm, 50 mW, Kimmon, Japan) is coaxial with the flow and 2) the optical system allows the measurement of light-scattering indicatrix from an individual particles. For each single particle, the time-dependent light scattering signal, so called "native SFC trace", is obtained, while the particle is moving through the testing zone. The native SFC traces are transformed to indicatrices. SFC allows measurements of indicatrices at polar angle ranging from 5$^o$ to 100$^o$ with integration over the azimuthal angle from 0$^o$ to 360$^o$. The angular resolution of the indicatrix measurement is better than 0.5$^o$. Fluorescence excited by Ar laser irradiation could be measured simultaneously with light scattering. Accurate measurements of light scattering indicatrices and fluorescence are available with the rate of 300 particles per second. To simulate the experimental indicatrices, the Mie theory and Wentzel-Kramers-Brillouin (WKB) approximation were used [23].

*Kinetic Measurements of Latex Agglutination*

The initial stage of agglutination of the BSA coated latex particles mixed with anti-BSA antibodies was studied in the series of the experiments with different initial concentration of the latex particles ranging from $3.8 \cdot 10^7$·cm$^{-3}$ to $7.5 \cdot 10^7$ cm$^{-3}$. All experiments were carried out at room temperature (25$^o$C). Buffered saline (PBS, pH=7.4) was used as a solvent for all reagents. Prior to the experiments, latex particle suspension was sonicated in an ultrasonic bath to break up initial aggregates. The experiments were performed as follows. 150 ml of the anti-BSA antibodies dissolved in buffered saline were mixed with equal volume of the preliminary sonicated BSA-coated particles. Then, during 1.5 hours aliquots (1 μl) were taken away at different time points and dissolved in 50 μl of the PBS to stop agglutination reaction. Then the sample was measured with the Scanning Flow Cytometer. In order to examine this stopping procedure, a few samples were measured twice: immediately after stopping procedure and 2 hours later. There was no noticeable difference in these two data sets. As for the procedure of the sample preparation, the captured aliquot was very small in comparison with the reaction volume. Therefore, the influence of this operation was negligible.

**THEORY**
*Mathematical Model*

The following processes are possible in the system: binding of the antibodies, *Y*, from the solute to antigens, *A*, immobilized on latex particles, dissociation of the antigen-antibody complexes, $AY$, and agglutination caused by the linking of particles, $L_i$, with the immobilized antibodies to antigens anchored with another particle:

$$A + Y \underset{k_-}{\overset{k_+}{\Leftrightarrow}} AY \qquad (1)$$

$$L_i + L_j \overset{k}{\Rightarrow} L_{i+j} \qquad (2)$$

We assume that the rate of particle-particle binding is much less then the rate of antibody-antigen binding. This assumption is valid in a wide range of experimental conditions [1,10]. Therefore, the reaction (1) can be treated in equilibrium, and the amount of antigen-antibody complexes, $N_{AY}$, on a monomer particle is determined by the concentration of free antibodies in the medium, amount of free (not occupied) antigens on the particle, $N_A$, and the dissociation constant $K = k_- / k_+$:

$$N_{AY} = \frac{Y N_A}{K} \qquad (3)$$

and

$$Y = -\frac{1}{2}(I_0 - Y_0 + K) + \frac{1}{2}\sqrt{(I_0 - Y_0 + K)^2 + 4KY_0} \qquad (4)$$

where $Y_0$ is the initial concentration of antibodies, and $I_0$ is the maximum amount of antigen-antibody complexes which can be formed on latex surfaces per unit volume of the system. $I_0$ can be derived using the initial concentration of latex monomers, $n_1(0)$, and the maximum amount of antigen-antibody complexes, $N_{max}$, which can be formed on a single monomer particle: $I_0 = n_1(0) N_{max}$.

We consider the reaction (2) as an irreversible process. In is known that the evolution of aggregates can be properly described by a von Smoluchowski kinetics:

$$\frac{dn_i}{dt} = -\sum_{j=1}^{\infty} k(i,j) n_i n_j + \frac{1}{2}\sum_{j=1}^{i-1} k(j, i-j) n_j n_{i-j} \qquad (5)$$

where $n_i$ is the concentration of clusters containing $i$ particles, and $k(i,j)$ is the rate kernel. From the comparison of the Eq. (5) and the mathematical formalism of chemical kinetics, the following relationship can be established between the deterministic second-order rate constant $k_c(i,j)$ and the rate kernel $k(i,j)$:

$$k(i,j) = (1 + \delta_{ij}) k_c(i,j) \qquad (6)$$

where $\delta_{ij}$ is Kronecker symbol.

To date, two distinct classes of kinetic aggregation have been investigated. One class is diffusion-limited aggregation [24-29] (DLA), which corresponds to a reaction occurring at each encounter between clusters. The other class is reaction-limited aggregation [30-33] (RLA), where the reaction rate is limited by the probability of forming a bond upon collision of two clusters.

It should be noted that immunoagglutination is the particular case of the agglutination phenomena by the following two main reasons. First, the binging is possible only between small specific sites – the antigen site on particle 1 and the antibody site on particle 2. Therefore, one can expect significant steric factor for immunoagglutination kinetics that is determined by the size and the number of binding sites of antigens and antibodies on a particle. Second, it is generally believed that the activation energy for the association reaction between antigen and antibody is rather low. Therefore, the reaction rate of immunoagglutination can be calculated in the diffusion-limited regime taking into account a steric factor, which reduces the DLA rate constant. The steric factor is significant, since specific binding sites of antigens and antibodies occupy rather small surface fraction of a latex particle.

Let us first consider the process of diffusion-limited binding of ligand molecules onto a cell surface. In that case, the following approximate expression was proposed for the rate kernel [34]:

$$k = \left(\frac{1-p}{4N_c b D_A} + \frac{1}{4\pi r D_A}\right)^{-1} \qquad (7)$$

where $D_A$ is the diffusion coefficient of ligand molecules, $N_c$ is the amount of cell receptors, $b$ is the radius of a circular binding site of the receptor, $p$ is the fraction of the cell surface covered by binding sites:

$$p = \frac{nb^2}{4r^2} \qquad (8)$$

It was tested that the Eq. (7) fits well the experimental data on the dynamics of cell distribution on the amount of antigen-antibody complexes in the reaction of binding the cell receptors with ligand molecules dissolved in the medium [35]. Namely, the diffusion-limited reaction is slowed significantly by a steric factor, which is determined by the size and number of surface binding sites per a cell. The size of binding site is considered to be much less then the size of the corresponding molecule (antigen or antibody), i.e. $p \ll 1$, and, therefore, one can rewrite Eq. (7) into the form

$$k = \left(\frac{\pi}{2N_c \sqrt{f}} + 1\right)^{-1} k_D \qquad (9)$$

where

$$k_D = 4\pi r D_A \qquad (10)$$

and $f$ is the surface fraction occupied by one binding site:

$$f = \frac{b^2}{4R^2} \qquad (11)$$

Now, let us consider the process of diffusion-limited binding of two latex particles, and each particle has only one binding site. If the surface fraction $f_A$ of the binding area of particle 1 is equal to the surface fraction $f_B$ of the binding area of particle 2, i.e. $f = f_A = f_B$, then [36,37]:

$$k_{12} = \left(\frac{1}{f^{3/2}} + 1\right)^{-1} k_D \qquad (12)$$

If the particles have many binding sites, the following modification of Eq. (12) can be suggested for the agglutination rate kernel, taking into account Eq. (9):

$$k_{12} = \left(\frac{1}{N_1 N_2 f^{3/2}} + 1\right)^{-1} k_D \qquad (13)$$

where $N_1$ and $N_2$ are the number of binding sites on the particles 1 and 2, correspondingly. Based on the Eq. (13), the following expression is assumed for the DLA rate kernel of immunoagglutination between clusters $i$ and $j$:

$$k(i,j) = \left(\frac{1}{ij\beta} + 1\right)^{-1} k_D(i,j) \qquad (14)$$

where
$$\beta = N_{1A} N_{1AY} f^{3/2} \tag{15}$$
here $N_{1A}$ and $N_{1AY}$ are the number of antigens and antibodies on a monomer latex particle, correspondingly.

In the case of the clusters with the fractal dimension ~ 3, the DLA coagulation kernel can be estimated approximately as [24] $k_D(i,j) \propto \left(i^{\frac{1}{3}} + j^{\frac{1}{3}}\right)\left(i^{-\frac{1}{3}} + j^{-\frac{1}{3}}\right)$, and, therefore, one can use the following estimation for $k_D(i,j)$:

$$k_D(i,j) = \frac{1}{2}\left(i^{\frac{1}{3}} + j^{\frac{1}{3}}\right)\left(i^{-\frac{1}{3}} + j^{-\frac{1}{3}}\right) k_D(1,1) \tag{16}$$

where
$$k_D(1,1) = \frac{4k_B T}{3\eta} \tag{17}$$

here $T$ is temperature, $k_B$ is Boltzmann constant, $\eta$ is the viscosity of the medium. For example, in water at room temperature ($T = 298°$ K): $k_D(1,1)=5.5 \cdot 10^{-12}$ cm$^3$s$^{-1}$.

Assuming Eq. (17) for $k_D(1,1)$ and Eq. (15) for $\beta$, Eq. (14) was used in this work in order to model the kinetics of latex immunoagglutination at the initial stage. The initial stage is considered as a stage when only monomers, dimers and trimers are taking into account, and the concentrations of higher multimers are negligible:

$$\begin{aligned}
\frac{dn_1}{dt} &= -k(1,1)n_1^2 - k(1,2)n_1 n_2 - k(1,3)n_1 n_3 \\
\frac{dn_2}{dt} &= \frac{1}{2}k(1,1)n_1^2 - k(1,2)n_1 n_2 - k(2,2)n_2^2 - k(2,3)n_2 n_3 \\
\frac{dn_3}{dt} &= k(1,2)n_1 n_2 - k(1,3)n_1 n_3 - k(2,3)n_2 n_3 - k(3,3)n_3^2
\end{aligned} \tag{18}$$

The system of Equations (18) does not have analytical solution in general case. Therefore, the evolution of the agglutination process was calculated from Eqs. (18) numerically.

**RESULTS and DISCUSSION**
***Light Scattering Indicatrices Detected at the Initial Stage of Agglutination.***
In order to verify the proposed method and to demonstrate the potential of the SFC in discrimination of clusters of spheres, the preliminary non-kinetic experiments with uncovered fluorescent latex particles were carried out. The particle sample was not sonicated. A non-specific aggregation of particles, resulted in formation of clusters of two and more particles, occurred in these conditions. Thus, single spheres ("monomers"), clusters of two spheres ("dimers"), and so forth were observed in measured sample. Both light-scattering indicatrix and fluorescence signals were recorded simultaneously for each measured particle.

The experimental signals measured by the Scanning Flow Cytometer from different particles are presented in Fig.1. The SFC signal consists of three separate parts. The first part is the light-scattering trace from a latex particle. The second part is the trigger pulse, which occurs when the particle passes a certain place (trigger position) in the testing zone. The third part is the fluorescence signal. The intensity of the fluorescence pulse is proportional to the number of dye molecules in measured particle. Thus, the intensity of fluorescence signal can be used to determine dimers, trimers and other clusters of fluorescent particles. The light scattering indicatrices evaluated from the corresponding FSC traces of the corresponding particles are presented in Fig. 2. One can see from Fig. 1 and Fig. 2, that light-scattering indicatrices can be used for monomer, dimer and higher multimers identification.

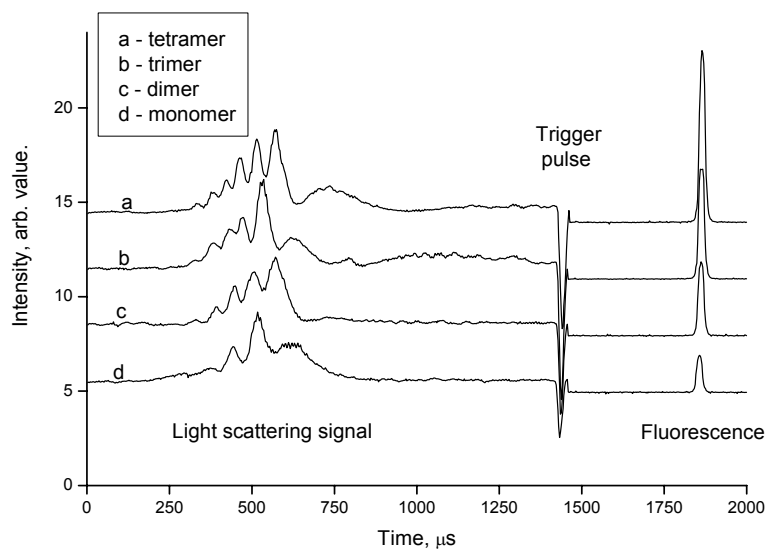

Fig. 1. Native SFC records of single particles from the sample with uncovered non-sonicated fluorescent latex particles. The signals are shifted on intensity for convenient view. Each signal consists of three parts (from left to right on the picture): light scattering signal, trigger pulse, and fluorescence signal from particle. Number of spheres in a cluster was determined with fluorescence intensity.

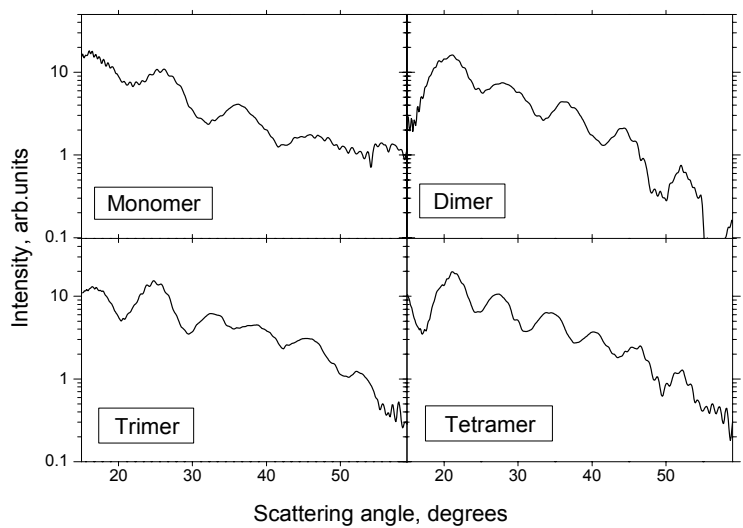

Fig. 2. Light-scattering indicatrices for the same particles as depicted in Fig. 1. Native SFC records from Fig. 1. were transformed into angular dependency using the transfer function.

The size and refractive index of homogeneous spherical particles are evaluated from light-scattering indicatrix using the parametric solution of the inverse light scattering problem. The particle characteristics obtained from measured indicatrices are shown in Fig. 3a (refractive index vs. size) and Fig. 3b (fluorescence vs. size). In Fig. 3 each point corresponds to one particle with its fluorescence, size and refractive index. It should be noted that the parameterization used in this work can be applied only for spherical particles, and it gives an effective size and refractive index for dimers. The particles were attributed to two groups: the left group consists of single spheres confirmed by their size (1.8 μm that is in agreement with manufacturer specification), and the right group corresponds to the clusters of two spheres. Good correlation between the obtained size and the fluorescence intensity leads to the conclusion that SFC is able to discriminate dimers from monomers from light scattering indicatrix.

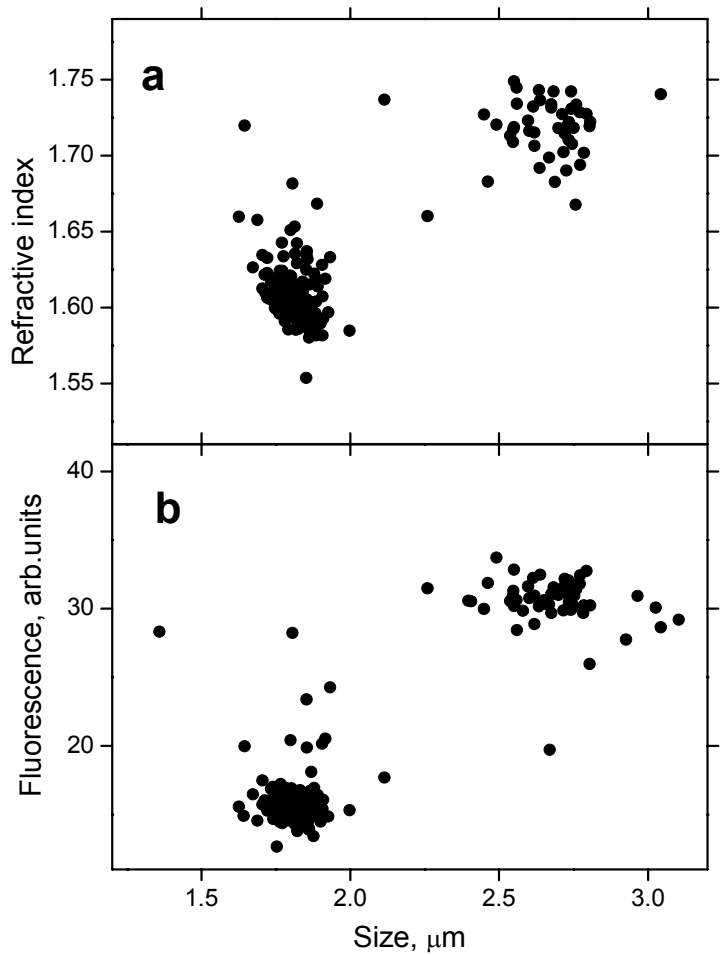

Fig. 3. Parameters of particles from the sample containing non-sonicated fluorescent latex spheres. Each point represents a single measured particle: A) refractive index vs. size map, the parameters were obtained from light scattering indicatrix with the FLSI-method; B) fluorescence vs. size map.

These experiments have shown that SFC allows the discrimination of "monomers" and "dimers" from light-scattering indicatrices, and, therefore, the SCF can be used for the study of the kinetics of the initial stage of latex agglutination.

### Kinetics of the initial stages of agglutination

Measurements of the kinetics of the immunoagglutination process initiated by the mixing of the BSA covered latex particles with anti-BSA IgG were performed during ~ 90 minutes from the beginning of the reaction. The experimental data obtained with different initial concentrations of latex particles and the same concentration of antibodies is presented in Figs. 4 - 6. Points represent the experimental data on monomers (Fig. 4), dimers (Fig. 5) and higher multimers (Fig. 6) fractions. The experimental data were fitted with Eqs. (18). The solid lines shown in Figs. 4 - 6 are the best fit to the experimental data. For every data set (i. e. independently for evolution of monomers, dimers and higher multimers at different initial concentrations of latex particles) the correspondent value of the parameter $\beta$ was obtained. Based on these data, the mean value of $\beta$ was evaluated: $\overline{\beta} = 0.21 \pm 0.03$ (the error includes also the variation of $\beta$ for every data set). The variation of $\beta$ for different data set was negligible (i.e. within its error).

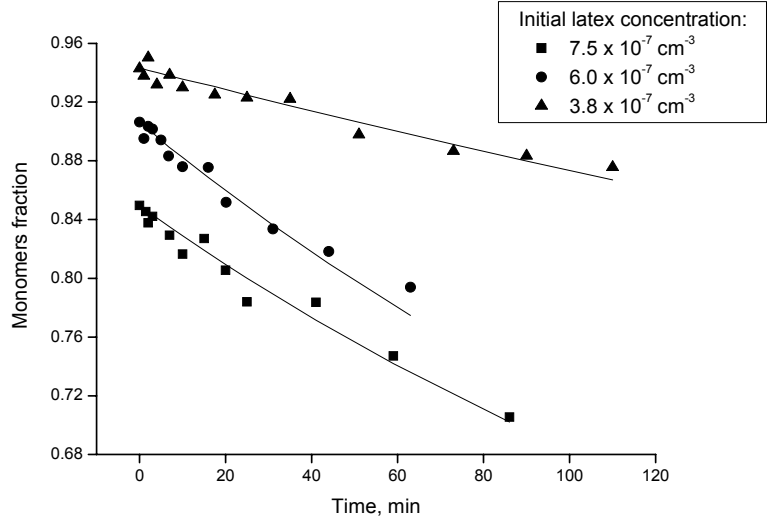

Fig. 4. Kinetics of the monomers fraction decrease at the initial stage of the agglutination process for BSA covered latex particles initiated by adding anti-BSA IgG: dots – experiment; solid line - theory. Initial concentration of anti-BSA is 0.8 μM.

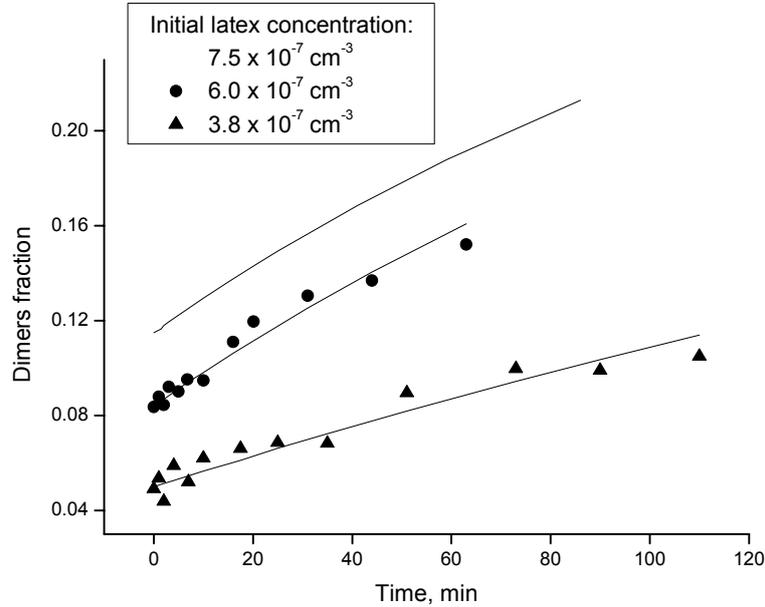

Fig. 5. Kinetics of the dimers fraction growth at the initial stage of the agglutination process for BSA covered latex particles initiated by adding anti-BSA IgG: dots – experiment; solid line - theory. Initial concentration of anti-BSA is 0.8 μM.

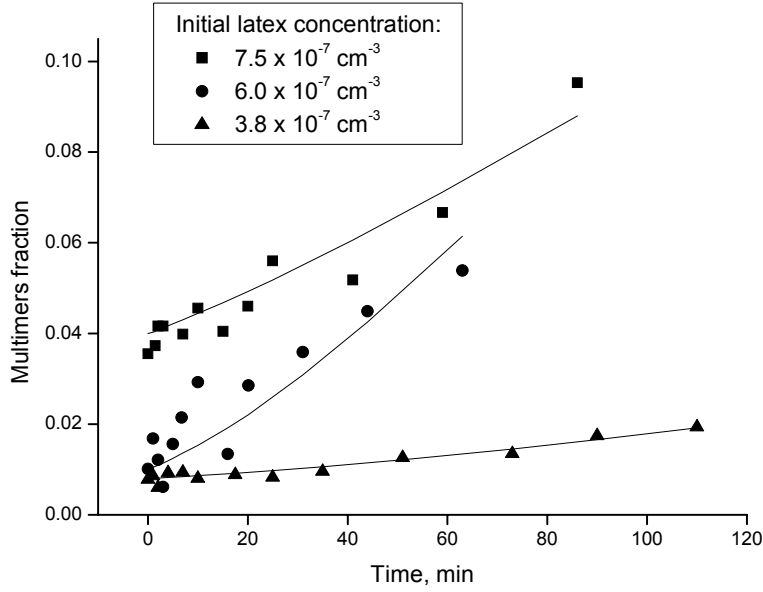

Fig. 6. Kinetics of the multimers (trimers and higher multimers) fraction growth at the initial stage of the agglutination process for BSA covered latex particles initiated by adding anti-BSA IgG: dots – experiment; solid line - theory. Initial concentration of anti-BSA is 0.8 μM.

The obtained value for mean $\beta$ was analyzed from the point of view of the known geometrical parameters of the antigen and the antibody. Hydrodynamic dimension of BSA [38] is $14 \times 3.8 \times 3.8$ nm$^3$. BSA can adsorb onto latex particles by two modes: side-on and end-on. In general, BSA is first adsorbed by side-on mode, followed by conformational change into end-on mode [39-40]. Immunoglobulin G molecules (IgG) are Y-shape assemblies of three rods (about 10 nm length and 4 nm thickness) joined in a fairly flexible region. Each IgG molecule is endowed with two identical antigen-specific binding sites. A typical binding site may be viewed [41-42] as a cleft of variable depth (0.5–1 nm), 1.5–2 nm length and about 1 nm width, as determined with x-ray crystallography. Based on this information and the obtained value of $\beta$, one can estimate other parameters of the immunoagglutination process, as follows. The product $N_{1A} N_{1AY}$ is estimated from Eq. (15) as

$$N_{1A} N_{1AY} = \frac{\beta}{f^{3/2}} = \frac{8R^3 \beta}{b^3} \qquad (19)$$

Taking the value of ~ 1 nm for the radius $b$ of binding site, one can get: $N_{1A}N_{1AY} \approx 1.2 \times 10^9$. On the other hand, it follows from Eq. (3) that

$$N_{1A} N_{1AY} = \frac{K/Y}{(1+K/Y)^2} N_{max}^2 \qquad (20)$$

where

$$N_{max} = \frac{4R^2}{d_{AY}^2} \qquad (21)$$

here $d_{AY}$ is the radius of the antigen-antibody complex. Taking the value of ~ 4 nm for $d_{AY}$, one can obtain $N_{max} \approx 2 \times 10^5$, and

$$\frac{K/Y}{(1+K/Y)^2} = \frac{N_{1A}N_{1AY}}{N_{max}^2} \approx 3 \times 10^{-2} \qquad (22)$$

From Eq. (22), taking into account the known concentration Y of antibodies in the medium, we get $K \approx 2 \times 10^{-8}$ M. This value corresponds to the typical dissociation constants for the antigen-antibody complex. From practical point of view, Eqs. (19)-(21) can be applied another way, for example: the known value of $K$ is used to evaluate unknown concentration $Y$ of antibody in the medium from the initial stage kinetics of immunoagglutination.

**CONCLUSIONS**

To sum up, the presented work demonstrates the advantage of the Scanning Flow Cytometer for the study of the kinetics of the formation of non-spherical particles in the process of latex immunoagglutination. The method proposed was verified by simultaneous measurements of light-scattering indicatrices and fluorescence from single particles. Kinetic measurements for the initial stage of latex agglutination initiated by mixing BSA covered particles and anti-BSA antibodies were carried out. The proposed mathematical model is based on the diffusion-limited approach with a steric factor. The steric factor is determined by the size and the number of binding sites on the surface of a latex particle. The obtained data are in good agreement with the mathematical modeling. The authors believe that the experimental method and the mathematical model open a new facility for quantitative determination of antibody concentrations in serum and for other biological and medical application.

**ACKNOWLEDGMENTS**

This research was financially supported by the following grants: grant of the Siberian Branch of the Russian Academy of Sciences (No. 115-2003-03-06); grants of the Russian Foundation for Basic Research (No. 02-02-08120 and No. 03-04-48852-a); NATO Science for Peace project SfP-977976. The authors thank Dr. G.M. Ignatiev (SRC VB VECTOR) for kindly providing anti-BSA antibodies, and Dr. B.A. Burmistrov (VECTOR-BEST) and Prof. V.B. Loktev (SRC VB VECTOR) for their support of this work. The authors express their gratitude to Prof. A.B. Doktorov (ICKC) and Prof. P.A. Purtov (ICKC) for valuable discussions.